\documentclass[preprint,showpacs,preprintnumbers,amsmath,amssymb]{revtex4}
\usepackage[dvips]{graphicx}
\usepackage{graphicx}
\usepackage{amsfonts}
\usepackage{bm}
\usepackage{amsmath}
\usepackage{amssymb}
\usepackage{color}
\usepackage[all]{xy}

\def\be{\begin{equation}}
\def\ee{\end{equation}}
\def\bea{\begin{eqnarray}}
\def\eea{\end{eqnarray}}

\begin{document}
\title{Black shells and naked shells}

\author{Walter Pulido}
\affiliation{Departamento de F\'\i sica, Universidad Nacional de Colombia, Bogot\'a, Colombia}
\email{wapulidog@unal.edu.co}

\author{Hernando Quevedo}
\email{quevedo@nucleares.unam.mx} 
\affiliation{Instituto de Ciencias Nucleares, Universidad Nacional Aut\'onoma de M\'exico, AP 70543, Ciudad de M\'exico 04510, Mexico}
\affiliation{Dipartimento di Fisica and ICRA, Universit\`a di Roma "Sapienza", I-00185, Roma, Italy}
\affiliation{Institute of Experimental and Theoretical Physics, 
	Al-Farabi Kazakh National University, Almaty 050040, Kazakhstan}

\date{\today}

\begin{abstract}

We study the collapse of a thin dust shell from the point of view of the horizon dynamics.  We identify the critical surfaces at which time and space coordinates interchange their roles and investigate their properties by using the formalism of trapped surfaces. We show the existence of marginally outer trapped surfaces that are associated with the presence of quasi-local horizons. A particular shell configuration that avoids the formation of horizons is interpreted as  naked shell.

\end{abstract}

\maketitle

\section{Introduction}
\label{sec:int}

The gravitational collapse is one of the most interesting predictions of general relativity. It is associated with the formation of black holes and gravitational waves, which are expected to contain information about the end state of highly interacting compact objects and about the dynamics of the physical processes that occur during the collapse. To find out the details of the formation of black holes and gravitational waves in general relativity, it is necessary to consider the entire set of Einstein equations and apply several methods of numerical relativity to find numerical solutions. Numerical relativity is a research area by itself that implies the use of highly accurate computational tools \cite{alcubi}. 

An alternative approach consists in considering only the most essential aspects of the gravitational collapse by analyzing an idealized model that reduces the complexity of the problem. This is the case of the black shell scenario,  a toy model in which a thin shell made of matter collapses under the influence of its own gravitational field \cite{isr66,poisson09,shells}. In this case, the mathematical complexity of the problem reduces drastically and, as a consequence, we are  allowed to apply mainly analytical methods. In the black shell model, we will assume that the contraction of a spherically symmetric shell starts at some radial distance and leads to a reduction of the shell radius with respect to a fiducial observer located at infinity. As the shell shrinks, the evolution is assumed to be described by an Oppenheimer-Snyder collapsing process \cite{oppen}. 

In this work, to analyze the dynamics of the  surface, where the thin shell is located, we consider the norm of the vector orthogonal to the surface and investigate the conditions under which critical surfaces appear, i.e., radii at which the norm of the vector vanishes and an interchange between time and space coordinates occurs. 
This method allows us to find all the critical surfaces that can appear during the evolution of the shell. For instance, this procedure predicts the existence of a horizon that appears as the shell radius equals its gravitational radius.  Furthermore, we will see that, in general, there exists a second critical surface with a radius that is always greater than the gravitational radius of the shell. To investigate the properties of the critical surfaces, we use the formalism of trapped surfaces and quasi-local horizons. We show that once the shell reaches its gravitational radius, space and time coordinates interchange their roles and the corresponding surface is a null surface with zero expansion so that it can be interpreted as an apparent horizon. The second surface corresponds to a marginally outer trapped surface that can be associated with a quasi-local horizon. 

We will see that there exists a particular case in which no horizons appear during the evolution of a shell, whose end state corresponds to a curvature singularity. We call this particular configuration a naked shell. Some properties of naked shells are also studied.

This paper is organized as follows. In Sec. \ref{sec:dyn}, we review the main aspects of the dynamics of a thin shell by using the Darmois-Israel formalism. 
We limit ourselves to the case of a spherically symmetric thin shell made of dust so that the corresponding equation of motion reduces to a first-order ordinary differential equation that turns out to be integrable. In Sec. \ref{sec:black}, we perform a detailed analysis of the critical surfaces that appear during the collapse of a shell. For a particular shell it is shown that no critical surfaces appear that could lead to the formation of horizons. This is the case of naked shells which is described in Sec. \ref{sec:naked}. 
The properties of the critical surfaces are described in Sec. \ref{sec:trapped} by investigating the behavior of the expansion of null vectors orthogonal to the surfaces. It is established that the critical surfaces correspond to marginally outer trapped surfaces.  
Finally, in Sec. \ref{sec:con}, we review the main results of our work and comment on possible future tasks to be investigated.

\section{Dynamics}
\label{sec:dyn}

In this section, we will follow the Darmois-Israel formalism \cite{isr66,isr67,hoye, nqs98,salim,dms20} in which the starting point is a spherically symmetric thin shell described by the hypersurface $\Sigma$ with coordinates $\xi^a=\{\tau,\theta,\varphi\}$. The corresponding line element on $\Sigma$ is assumed to be of the form
\be
ds^2_\Sigma = - d\tau^2 + R^2(\tau)d\Omega^2\ .
\label{shellmet}
\ee
 Thus, $\Sigma$ splits the spacetime into two regions $V_-$, inside $\Sigma$, and $V_+$, outside $\Sigma$. The thin shell is assumed to be described by an energy-momentum tensor $S^{ab}$.  

To describe the spacetime, we assume that the inside region $V_-$ corresponds to the Minkowski spacetime 
\begin{equation}\label{302}
ds^2_{-}=-dt^2+dr^2+r^2d\Omega^2 \ .
\end{equation}
On $\Sigma$, it is convenient to introduce new coordinates ${T}(\tau)$ and $R(\tau)$ such that 
\be
t = {T}(\tau)\ , \  dt=\dot{{T}}d\tau\ ,\quad 
r=R(\tau)\ ,\ dr=\dot{R}d\tau\ ,
\label{coord1}
\ee 
where $\tau$ is the proper time and a dot represents derivative with respect to $\tau$. Thus, the interior Minkowski metric on $\Sigma$ becomes 
\begin{equation}\label{303}
ds^2_{-}|_\Sigma=-\Bigl(\dot{{T}}^2-\dot{R}^2\Bigr)d\tau^2+R^2(\tau)d\Omega^2\ .
\end{equation}

Furthermore, we will assume that the outside region corresponds to the Schwarzschild spacetime
\begin{equation}\label{307}
ds^2_{+}=-f dt^2+\frac{dr^2}{f }+r^2d\Omega^2\ ,\ f = 1 - \frac{2M}{r}, 
\end{equation}
which on $\Sigma$ in coordinates (\ref{coord1}) becomes
\begin{equation}\label{308}
ds^2_{+}|_\Sigma=-\Bigl(F\dot{T}^2-\frac{\dot{R}^2}{F}\Bigr)d\tau^2+R^2(\tau)d\Omega^2,
\ F = 1 - \frac{2M}{R}\ .
\end{equation}

To guarantee that the entire spacetime is well defined as a differential manifold, one can impose the Darmois matching conditions 
\be
[h_{ab}] = h_{ab}^+ - h_{ab}^- =0 \ , \quad [K_{ab}] =K_{ab}^+ - K_{ab}^-=0\ ,
\label{darmois}
\ee
where $h_{ab}^\pm$ is the metric induced on $\Sigma$ by the metric of $V_\pm$ and
$K_{ab}^\pm$ is the corresponding extrinsic curvature, respectively. The first condition implies simply that $ds^2_{+}|_\Sigma=ds^2_{-}|_\Sigma$, i.e., 
\be
\dot T^2 - \dot R^2 = F\dot{T}^2-\frac{\dot{R}^2}{F}\ .
\ee

In general, it is quite difficult to satisfy the second condition of Eq.(\ref{darmois}). A less strict version of this condition was proposed by Israel and consists in assuming that the jump in the extrinsic curvature, $[K_{ab}] \neq 0$, determines a thin shell with energy momentum tensor $S_{ab}$, which is defined as
\be
S_{ab} = - \frac{1}{8\pi} ( [K_{ab}]- [K] h_{ab})
\label{emcond}
\ee
where $K = K_{ab} h^{ab}$. 
For simplicity, let us consider the case of a dust shell $S_{ab}= \sigma u_a u_b$, where $\sigma$ is the surface density of the dust and $u_a$ is the 3-velocity of the shell. It is then straightforward to compute the extrinsic curvature of $V_+$ and $V_-$ and the right-hand side of Eq.(\ref{emcond}), which determines the behavior of the surface density $\sigma$. The final result can be expressed as
\be
R(\sqrt{1+\dot R ^2} - \sqrt{F+\dot R ^2})  = m = 4\pi \sigma R^2 
\label{moteq1}
\ee 
where $m$ is an integration constant. This equation can be interpreted as the motion equation of the shell. Indeed, a rearrangement of Eq.(\ref{moteq1}) leads to the expression
\be
\label{moteq2}
M = m \sqrt{1+\dot R ^2} - \frac{m^2}{2R} \ ,
\ee
which can be interpreted as representing the conservation of energy during the motion of the shell. In fact, the first term on the right-hand side represents a  relativistic quantity, which includes the energy at rest and the kinetic energy. Then, the second term can be interpreted as the binding energy of the system.
Consequently, $M$ represents the gravitational mass of the shell and $m$ its rest mass \cite{poisson09}.  The equation of motion (\ref{moteq2}) can be rewritten as 
\be
\dot R^2 = \left(\frac{M}{m} + \frac{m}{2R} -1\right)\left(\frac{M}{m} + \frac{m}{2R} +1\right)\ .
\label{moteq3}
\ee
Since the right-hand side of this equation must be positive, it follows that if $m\geq M$ 
 the radius of the shell can take values only within the interval
\be
R\in \Big(0, \frac{m^2}{2m-2M} \Big]
\ee
with boundaries 
\bea
{\rm for}\ m\rightarrow M & \Rightarrow & R \in (0,\infty) \ ,\nonumber \\
{\rm for}\ m = 2 M        & \Rightarrow & R \in (0,2M] \ , \\
{\rm for}\ m\rightarrow \infty & \Rightarrow & R \in (0,\infty)\ . \nonumber 
\label{interval}
\eea
On the other hand, if $m< M$, during the evolution of the shell its radius can have any positive value, $R\in (0,\infty)$. 
We see that the value of the rest mass $m$ is important for determining the motion of the shell. The lower limit $(R\rightarrow 0)$ follows from the interpretation of the function $R(\tau)$ as the radius of the shell and also, as we will show below,  from the fact that it corresponds to a curvature singularity.

\begin{figure}	
	\includegraphics[scale=0.3]{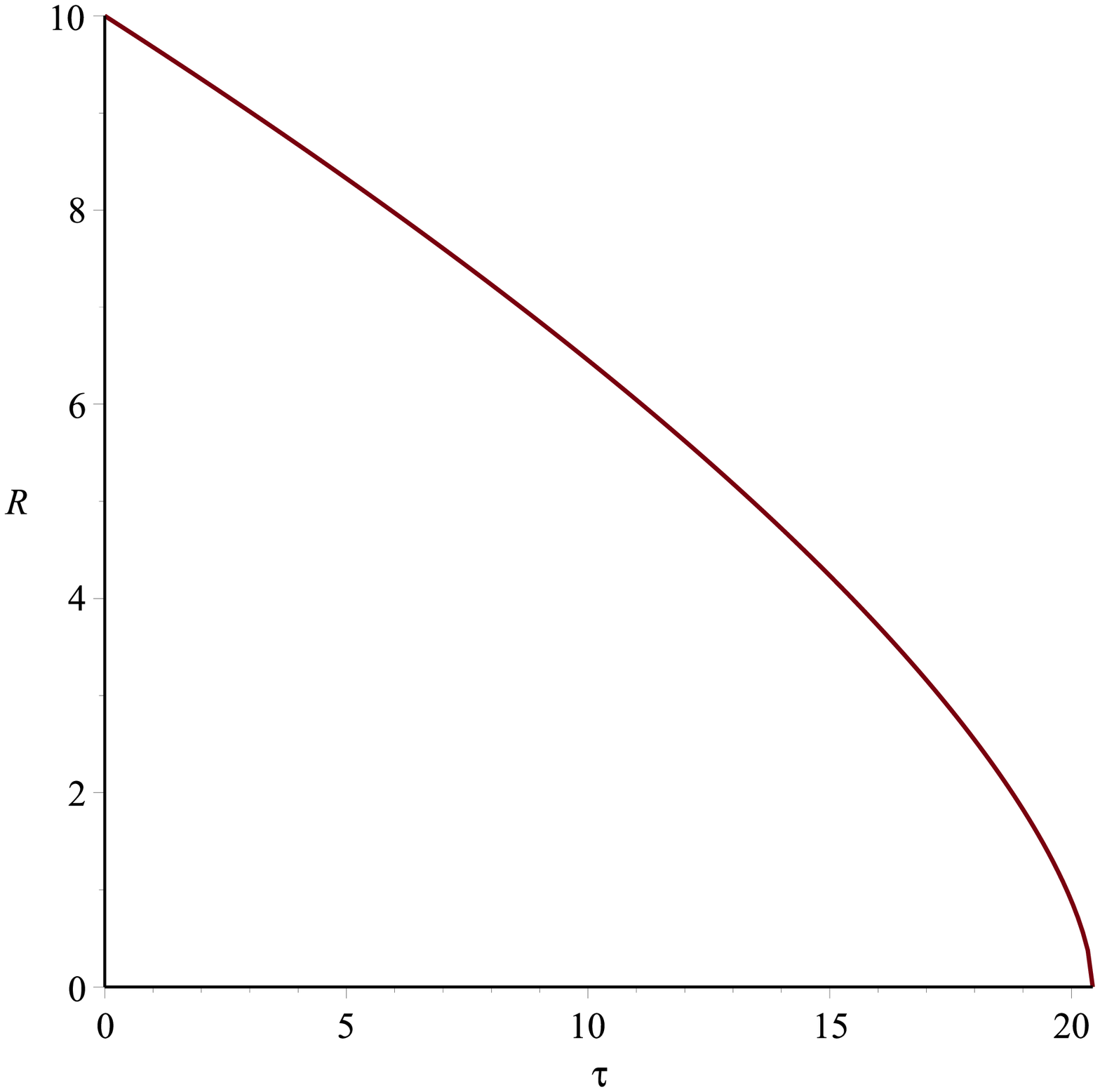}
	\includegraphics[scale=0.3]{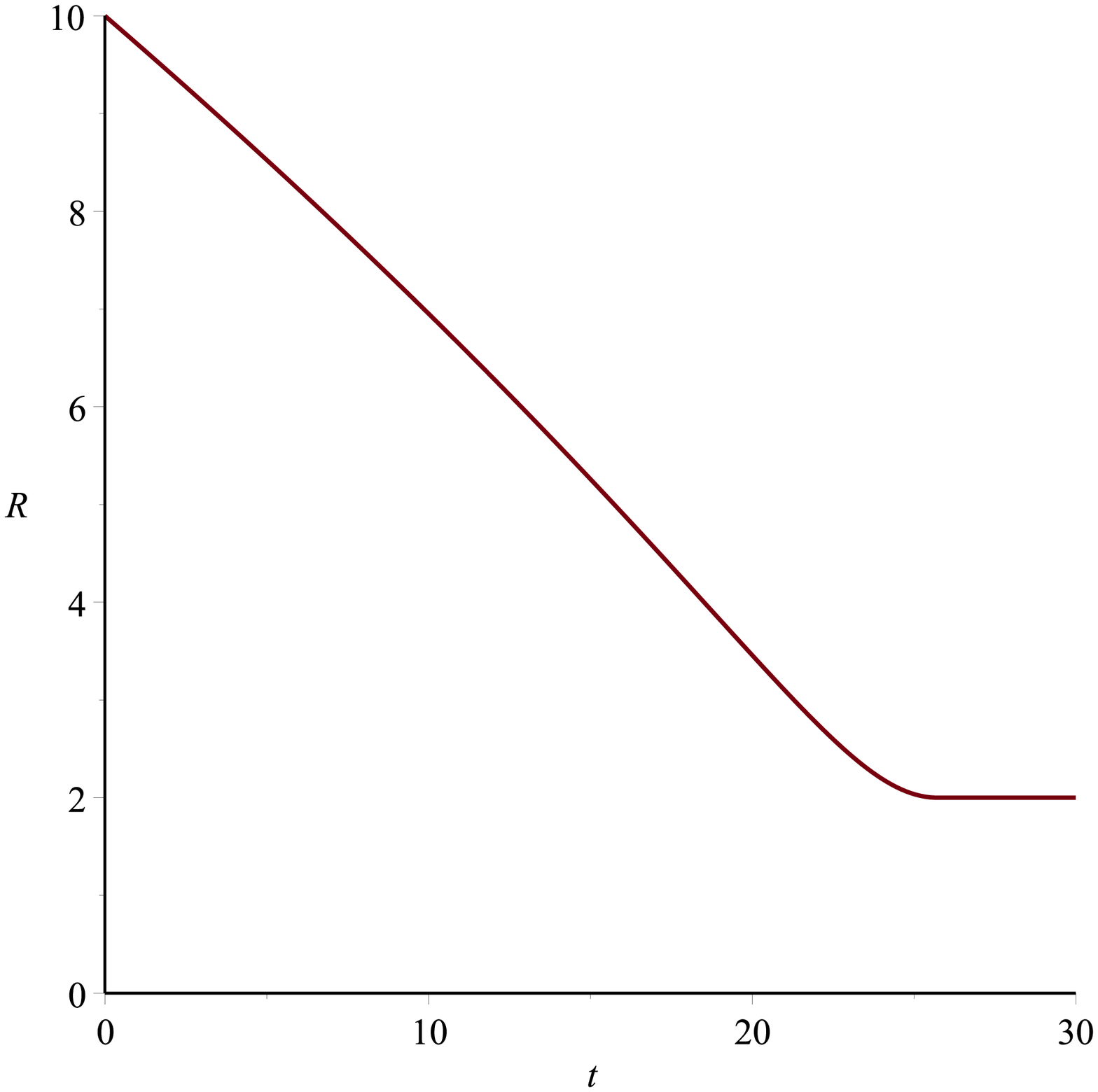}
	\caption{The radius of the shell in terms of the proper time $\tau$ and the coordinate time $t$ for the particular masses $M=1$ and $m=1$. \label{fig1} 	}
\end{figure}

The result of integrating the motion equation (\ref{moteq2}) is shown in Fig. \ref{fig1}. We present the result in terms of the proper time $\tau$ and the coordinate time $t$. As expected, in terms of the proper time $\tau$, the shell reaches the origin of coordinate in finite time, whereas for an observer at infinity the shell never reaches the radius $R=2M$. 

\section{Critical surfaces and horizons}
\label{sec:black}

In a spacetime, horizons are usually defined in terms of Killing vectors. For instance, if the spacetime is static with a timelike Killing vector $\xi^\mu$, the condition $\xi^\mu \xi_\mu =0$ determines a hypersurface which is interpreted as the event horizon. In the case of the shell we are considering here, the corresponding spacetime has no timelike Killing vector and so it is not possible to use the above definition to search for horizons. 
In general, in the case of time-dependent spacetimes, horizons must be treated in a different manner. We will see this in the next section. 

In this section, we will search for critical surfaces that resemble the properties of a horizon. To this end, let us consider a radially directed vector  
$U^\mu = (\dot T, \dot R, 0, 0)$ with norm
\be
U^2 = - F \dot T^2 + \frac{\dot R^2}{F}\ .
\label{norm0}
\ee
The behavior of the norm $U^2$ along the radial coordinate contains information about the surfaces orthogonal to it. For instance, the locations at which the norm vanishes can be interpreted as indicating the presence of a null surface, which is one of the properties of horizons. To be more specific, let us consider the case in which the norm  is timelike when the shell is at rest $\dot R =0$, i.e., $ F \dot T^2 = 1$. 
 This is a pure coordinate condition that determines how the time coordinate $T$ depends on the spatial coordinate $R$. For simplicity, we assume that this condition is valid all the way along the radial coordinate and so  the norm becomes
\be
U^2 = - 1 + \frac{\dot R ^2}{F}\ ,
\ee
which can be expressed as
\be
U^2 = \left[ - 2 + \frac{2M}{R} + \left(\frac{M}{m} + \frac{m}{2R}\right)^2\right] \left( 1- \frac{2M}{R}\right)^{-1} \ .
\label{norm}
\ee

\begin{figure}[ht]
	\includegraphics[scale=0.3]{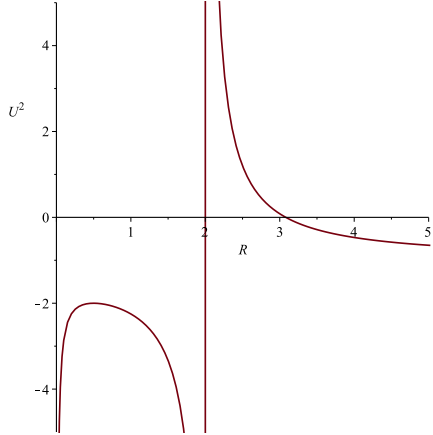}
	\caption{Norm of the observer's 4-velocity, according to Eq.(\ref{norm}) for the particular masses $M=1$ and $m=1$. \label{fig4} 	}
\end{figure}

In Fig. \ref{fig4}, we illustrate the behavior of $U^2$. We see that there are two critical points, where the norm vanishes, namely $R=R_{gr}=2M$ and $R=R_h$. The former value is obtained from the algebraic equation 
\be
4 (2m^2- M^2) R^2 -12 m^2 M R - m^4 = 0 \ ,
\ee
for which we find the positive solution 
\be
R_h  =  \frac{m^2 (3M+{\cal M})}{2(2m^2- M^2)} \ , \quad 
{\cal M}  =  \sqrt{2 m^2 + 8 M ^2}\ .
\label{hor}
\ee
In both cases, the surface is determined by the equation $R=const$ so that the corresponding  normal vector is $n_\mu = (0,1,0,0)$ with norm $n^2= g^{RR}= 1- 2M/R$. It follows that the surface $R=R_{gr} = 2M$ is the only null surface which,  according to the behavior of the norm $U^2$ around $R=2M$, resembles the properties of an event horizon.

\begin{figure}
	\includegraphics[scale=0.3]{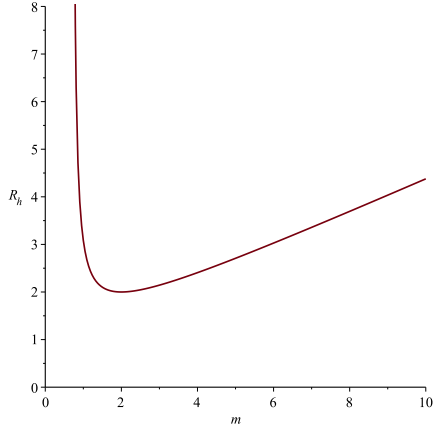}
	\caption{Location  of the horizon radius $R_h$ in terms of the mass $m$. Here we choose $M=1$, which means that $R_h$ and $m$ are given in multiples of $M$. 
		\label{fig3} 
	}
\end{figure}

The radius $R_h$ is given in Eq.(\ref{hor})  in terms of the gravitational mass $M$ and the rest mass $m$. The behavior of this quantity is illustrated in Fig. \ref{fig3}.
We can see two special points in this plot. First, for $m=M/\sqrt{2}$, the radius diverges, indicating that $R_h$ exists only for values of  $m>M/\sqrt{2}$. For rest masses with $m<M/\sqrt{2}$, Eq.(\ref{hor}) indicates that no radius exists $(R_h<0$). 
The second point, $R_h = 2M$ with $m=2M$, is a minimum value that corresponds to the Schwarzschild radius.  
In the particular case of a shell at rest at infinity 
$(\dot R=0, \ R\rightarrow\infty)$, from the equation of motion (\ref{moteq2}), it follows that the rest mass and the gravitational mass coincide, $m=M$, and then the equation for the radius $R_h$ reduces to 
\be
R_h = \frac{M}{2}(3 + \sqrt{10}) \ .
\label{radinf}
\ee

Figure \ref{fig4} shows the location of the radius $R_h$ in accordance with Eq.(\ref{hor}). Furthermore, we see that at $R=R_h$ the norm changes its sign.  This is an effect that is observed exactly on the horizon of black holes. 
Consequently, the  surface with radius $R_h$ denotes the particular location, where an interchange between the time and spatial coordinates takes place. However, this is not a null surface, which is a property of  event horizons. Nevertheless, we will see below that it can be associated with the presence of a quasi-local horizon.

According to the above results, a black shell consists of a central singularity with one or two horizons, which are located as follows: 
\bea
\label{blackshells}
{\rm if}\ m< \frac{M}{\sqrt{2}} & \Rightarrow & {\hbox{an event horizon at }} R=2M\ , \\
{\rm if}\ m\geq \frac{M}{\sqrt{2}} {\rm \ and \ } m\neq 2M & \Rightarrow & {\hbox{an event horizon  at }} R=2M\ {\hbox{and a dynamic horizon at}}\  R=R_h \ .\nonumber
\eea

The inner horizon located at $R=2M$ is always present, except in the case $m=2M$ that we will consider below.  The outer  horizon located at $R=R_h>2M$ is not always present; its existence and location depend on the value of the rest mass $m$. From Eq.(\ref{hor})  it follows that for the particular value $m=2M$, the radius of the exterior dynamic horizon $R_h$ reduces to its minimum value $R_h=2M$, i.e., it coincides with the inner horizon located at the Schwarzschild radius $R_S=2M$. This could be interpreted as a degenerate case in which the two horizons coincide. However, a detailed analysis shows that in this case, no horizon exists. This will be shown in the next section.

\section{Naked shells}
\label{sec:naked}

In this section, we will consider a particular configuration that can exist only for a very specific value of the rest and gravitational masses. 
From the expression for the dynamic horizon radius given in Eq.(\ref{hor}), it follows that for the particular value $m=2M$, the radius  $R_h$ reduces to its minimum value $R_h=2M$, i.e., it coincides with the  horizon located at the Schwarzschild radius $R_S=2M$. This could be interpreted as the degenerate case in which the two horizons coincide. 

However, a straightforward computation of the norm $U^2$ leads to the expression 
\be
U^2(m=2M)= -\frac{2M+7R}{4R} \ ,
\ee
which has no zeros for any positive values of $R$. This means that during the evolution of a particular shell, in which the rest mass is twice the gravitational mass, no horizons are formed. Moreover, the end state of the shell evolution corresponds to a curvature singularity. 
Indeed, the computation of the Kretschmann scalar for the shell metric (\ref{shellmet}) leads to the expression 
\be
K = R_{abcd} R^ {abcd} = 4
\frac{ 1 + 2 \dot R ^2 + \dot R ^4 + 2 \ddot{R} R^2 }{R^4} \ .
\ee	
We see that the only singularity occurs when $R\rightarrow 0$, i.e., as the radius of the shell shrinks to its minimum value.  No other singularities exist during the collapse of the shell as long as its velocity and acceleration remain finite. 

The particular configuration described above in which a curvature singularity is formed as the end state of evolution of a thin shell, but no horizons appear during the evolution, will be called naked shell. It exists only for a very specific value of the rest mass. For any other value of the rest mass, the collapse of the shell is characterized by the appearance of horizons, implying that the corresponding configuration is a black shell.

From the equation of motion (\ref{moteq3}), it follows that in the case of a naked shell the dynamics is governed by the equation
\be
\dot R = - \sqrt{\left(\frac{M}{R} - \frac{1}{2}\right)\left( \frac{M}{R} + \frac{3}{2} \right) }\ ,
\ee
where the minus sign has been chosen in order for the equation to describe the motion of a collapsing shell. The motion is constrained within the interval 
$R\in (0,2M]$. This means that a shell can start collapsing at any $R\leq 2M$, where $R=2M$ is not a horizon and will reach the singularity in a finite proper time. Any observer within the radial distance $R\leq 2M$ can communicate with an observer located infinitesimally close to the central singularity.

\section{Quasi-local horizons}
\label{sec:trapped}

As shown in Sec. \ref{sec:black}, during the collapse of a shell special surfaces can appear around which the time and spatial coordinates interchange their roles. The interpretation of these surfaces as event horizons is problematic because they are defined as global properties of spacetime. The physical evolution of a global horizon is at least problematic, if not impossible. For this reason, an alternative approach is necessary in which local properties of spacetime are invoked. This is the formalism of quasi-local horizons which is based upon the concept of trapped surfaces, dating back to the formulation of the singularity theorems proved by Penrose \cite{penrose65} and Hawking and Ellis \cite{hawellis73}. In turn, trapped surfaces are defined in terms of the behavior of the expansion of null vectors orthogonal to the surfaces. 

In the case under consideration, all the surfaces to be investigated are spherically symmetric, which allows us to perform all the calculations explicitly. Indeed, we are interested in the behavior of the critical surfaces described in Sec. \ref{sec:black}, which correspond to 2-spheres with $r=const.$
To be more specific, consider the null vectors $ l^\alpha$ and $N^\alpha$ with $l_\alpha N^\alpha= -1$, which are orthogonal to the surface $\Sigma$  of the shell. Then, the expansions of $l$ and $N$ are defined as \cite{skb06}
\be
\Theta_{(l)} = q^{\alpha\beta} \nabla_\alpha l_ \beta\ , \quad
 \Theta_{(N)} = q^{\alpha\beta} \nabla_\alpha N_ \beta\ , \label{expansions}
\ee
where the components of the metric $q^{\alpha\beta} $ are defined as
\be
q^{\alpha\beta} = g^{\alpha\beta} + l^\alpha N^\beta + N^\alpha l^\beta\ .
\label{metricq}
\ee
To determine the null vectors $l$ and $N$, we introduce the auxiliary orthonormal vectors $n^\alpha$ and $r^\alpha$, which are obtained by representing the spacetime metric $g^{\alpha\beta}$ in terms of the induced metric $h^{ab}$ as \cite{poisson09}
\be
g^{\alpha\beta} = n^\alpha n ^\beta + h^{ab} e_a^\alpha e_b ^\beta \ , 
\ee
with
\be
n_\alpha n^\alpha = 1\ , \quad
e_a^\alpha = \frac{\partial x^\alpha}{\partial \xi ^a }
\ee
where the induced metric $h_{ab}$ has been used to determine the line element on the surface of the shell $\Sigma$, i.e., 
\be
ds^2|_\Sigma = g_{\alpha\beta}dx^\alpha dx^\beta |_\Sigma = g_{\alpha\beta} e^\alpha_a e^\beta_ b d \xi^ d\xi ^b = h_{ab} d\xi ^a d\xi ^b\ .
\ee

A straightforward computation by using the exterior metric (\ref{307}) leads to the following expression for the components of the vector $n$
\be
n^\alpha = \frac{1}{(F\dot T ^2 - F^{-1} \dot R ^2 )^{1/2}}\left(\frac{\dot R}{F} , F\dot T , 0 , 0\right)\ .
\ee
Furthermore, we introduce a vector $r^\alpha$ such that $r^\alpha r_\alpha = -1 $ and $n_\alpha r^\alpha =0$. Then, we obtain 
\be
r^\alpha = 
\frac{1}{(F\dot T ^2 - F^{-1} \dot R ^2 )^{1/2}}\left(\dot T , \dot R, 0, 0\right)\ .
\ee
Finally, the null vectors are defined as 
\be
l^\alpha = \frac{1}{\sqrt{2}} (r^\alpha + n^\alpha )\ , \quad 
N^\alpha = \frac{1}{\sqrt{2}} (r^\alpha - n^\alpha )\ ,
\ee
 and can be written explicitly as
\be
l^\alpha = \frac{1}{\sqrt{2}(F\dot T ^2 - F^{-1} \dot R ^2 )^{1/2}}
\left(\frac{\dot R}{F}+  \dot T , F \dot T + \dot R , 0, 0\right)\ ,
\ee
\be 
N^\alpha = \frac{1}{\sqrt{2}(F\dot T ^2 - F^{-1} \dot R ^2 )^{1/2}}
\left(\dot T - \frac{\dot R}{F}, \dot R -F \dot T  , 0, 0\right)\ ,
\ee
which satsisfy the conditions $N_\alpha N^\alpha= l_\alpha l^\alpha = 0$ and $l_\alpha N^\alpha = -1$. These expressions can now be introduced into Eq.(\ref{metricq}) for the components of the metric $q$  to obtain $q_{\alpha\beta} = {\rm diag}(0,0,r^2,r^2\sin^2\theta)$. Then, the expansions (\ref{expansions}) can be computed and we obtain 
\be
\Theta_{(l)} = 0\ , \quad \Theta_{(N)} = 0\ ,
\ee
indicating that the surfaces with $r=const.$ are marginally outer trapped surfaces \cite{hayward94}. We notice that a similar computation by using the inner metric leads to the same result. 

We can now  identify the critical surfaces obtained in Sec. \ref{sec:black}. The surface $r=2M$ is thus a null surface with zero expansion and so it corresponds to an apparent horizon \cite{melia18}. On the other hand, the surface $r=R_h$, which  is not null but is accompanied by an interchange between the time and spatial coordinates, is a marginally outer trapped surface, which can be interpreted as corresponding to a quasi-local horizon.


\section{Conclusions}
\label{sec:con}

In this work, we analyzed the collapse of a spherically symmetric thin shell made of pure dust. To describe the dynamics of the shell, we employ the  Darmois-Israel formalism, according to which the complete spacetime is split into three different parts that must satisfy the matching conditions. In our case, the interior part corresponds to a flat Minkowski spacetime, the exterior one is described by the spherically symmetric Schwarzschild metric and the boundary between them is described by an induced metric that satisfies the matching conditions and can be interpreted as corresponding to a thin shell of dust. As a result of demanding compatibility between the three spacetime metrics, we obtain a differential  equation that governs the motion of the shell and  depends on the gravitational mass $M$ and on an additional integration constant $m$, which is interpreted as the rest mass of the shell.

By using a particular vector oriented along with the motion of the shell, we search for critical surfaces around which an interchange between space and time coordinates occurs, which resembles the behavior around a horizon. We found two critical surfaces.  The first one  appears when the radius of the shell equals its gravitational radius ($R=2M$), i.e., it corresponds to the Schwarzschild horizon $R_S$ of the exterior spacetime. The radius $R_h$ of the second critical surface is always greater than the gravitational radius of the shell and its explicit value depends on the values of the  gravitational and rest masses. 
The interpretation of these critical surfaces as horizons is problematic due to the dynamic behavior of the shell. The standard definition of an event horizon is associated with global properties of the spacetime, which makes it difficult or even impossible to analyze its evolution. Therefore, we use the alternative formalism of trapped surfaces, which is based on local properties of the spacetime. The idea is to analyze the behavior of null vectors orthogonal to the surface of the shell. It turns that all the surfaces with constant radial distance are characterized by a vanishing expansion. Consequently, the two critical surfaces located at $R_S$ and $R_h$ correspond to marginally outer trapped surfaces, which are associated with quasi-local horizons. In addition, since the critical surface located at the gravitational radius $R=2M$ is a null surface, it can be interpreted as an apparent horizon.

Furthermore, for the particular case of a thin shell, whose rest mass is twice the gravitational mass ($m=2M)$, no horizon exists and the end state of the shell evolution corresponds to a curvature singularity, which appears as the radius of the shell tends to zero. We thus denote the resulting configuration as a naked shell. This is a very  peculiar configuration that appears only because of the existence of the second critical surface $R_h$. Indeed, whereas the vector that we use to detect critical surfaces, for $m\neq 2M$ predicts the existence of two locations with zero norm at $R_S$ and $R_h$, in the limiting case with $m=2M$ shows no zeros at all. This means that during the collapse of a shell with 
$m=2M$ no interchange between space and time occurs and so no horizon appears that could 
hide the curvature singularity from the outer spacetime. 

We have shown that the critical  surfaces that appear during the collapse of a thin shell can be interpreted as quasi-local horizons, which imply the existence of a sophisticated structure from the point of view of dynamics. In a related context, the idea of interpreting black holes as macroscopic quantum objects was proposed in \cite{dms20}. The present work can be generalized to include another kind of thin shells and spacetimes. For instance, one could consider the case of thin shells with internal pressure or additional gravitational charges. We expect to study such generalized configurations in future works.

\begin{acknowledgments}
	This work was partially supported  by UNAM-DGAPA-PAPIIT, Grant No. 114520, 
	Conacyt-Mexico, Grant No. A1-S-31269, 
	and by the Ministry of Education and Science (MES) of the Republic of Kazakhstan (RK), Grant No. 
	BR05236730 and AP05133630. 
\end{acknowledgments}

\end{document}